# Anisotropic Magnetization Relaxation in Ferromagnetic GaMnAs Thin Films

**Kh.Khazen, H.J.von Bardeleben, M.Cubukcu, J.L.Cantin**
Institut des Nanosciences de Paris,
Université Paris 6, UMR 7588 au CNRS
140, rue de Lourmel, 75015 Paris, France

**V.Novak, K.Olejnik, M.Cukr**
Institut of Physics, Academy of Sciences,
Cukrovarnicka 10, 16253 Praha, Czech Republic

**L.Thevenard, A. Lemaître**
Laboratoire de Photonique et des Nanostructures, CNRS
Route de Nozay, 91460 Marcoussis, France

## Abstract:

The magnetic properties of annealed, epitaxial $Ga_{0.93}Mn_{0.07}As$ layers under tensile and compressive stress have been investigated by X-band (9GHz) and Q-band (35GHz) ferromagnetic resonance (FMR) spectroscopy. From the analysis of the linewidths of the uniform mode spectra the FMR Gilbert damping factor $\alpha$ has been determined. At T=4K we obtain a minimum damping factor of $\alpha = 0.003$ for the compressively stressed layer. Its value is not isotropic. It has a minimum value for the easy axes orientations of the magnetic field and increases with the measuring temperature. Its average value is for both type of films of the order of 0.01 in spite of strong differences in the inhomogeneous linewidth which vary between 20 Oe and 600 Oe for the layers grown on GaAs and GaInAs substrates respectively.

## Introduction:

The magnetic properties of ferromagnetic $Ga_{1-x}Mn_xAs$ thin films with Mn concentrations between x=0.03 and 0.08 have been studied in great detail in the recent years both theoretically and experimentally. For recent reviews see references [1, 2]. A particularity of GaMnAs ferromagnetic thin films as compared to conventional metal ferromagnetic thin films is the predominance of the magnetocrystalline anisotropy fields over the demagnetization fields. The strong anisotropy fields are not directly related to the crystal structure of GaMnAs but are induced by the lattice mismatch between the GaMnAs layers and the substrate material on which they are grown. When grown on (100) GaAs substrates the difference in the lattice constants induces biaxial strains of ≈ 0.2% which give rise to anisotropy fields of several $10^3$ Oe. The low value of the demagnetization fields (~300Oe) is the direct consequence of the small spin concentration in diluted magnetic semiconductors (DMS) which for a 5% Mn doping leads to a saturation magnetization of only 40 emu/cm$^3$. As the strain is related to the lattice mismatch it can be engineered by choosing different substrate materials. The two systems which have been investigated most often are (100) GaAs substrates and (100)GaInAs partially relaxed buffer layers. These two cases correspond to compressive and tensile strained GaMnAs layers respectively [3].

The static micro-magnetic properties of GaMnAs layers can be determined by magnetization, transport, magneto-optical and ferromagnetic resonance techniques. For the investigation of the magnetocrystalline anisotropies the ferromagnetic resonance spectroscopy (FMR) technique has been shown to be particularly well adapted [2, 4]. The dynamics and relaxation processes of the magnetization of such layers have hardly been investigated up to now [5-7]. The previous FMR studies on this subject concerned either unusually low doped GaMnAs layers [5, 7] or employed a single microwave frequency [6] which leads to an overestimation of the damping factor. The knowledge and control of the relaxation processes is in particular important for device applications as they determine for example the critical currents necessary for current induced magnetization switching. It is thus important to determine the damping factor for state of the art samples with high Curie temperatures of $T_C \approx$ 150K, such as those used in this work. Another motivation of this work is the search for a potential anisotropy of the magnetization relaxation in a diluted ferromagnetic semiconductor in which the magnetocrystalline anisotropies are strong and dominant over the demagnetization contribution.

The intrinsic small angle magnetization relaxation is generally described by one parameter, the Gilbert damping coefficient $\alpha$, which is defined by the Landau Lifshitz Gilbert (LLG) equation of motion for the magnetization:

$$\frac{1}{\gamma}\cdot\frac{d\vec{M}}{dt} = -\left[\vec{M}\times\vec{H}_{eff}\right] + \frac{\alpha}{\gamma}\left[\vec{M}\times\frac{d\vec{s}}{dt}\right] \qquad \text{eq.1}$$

with M the magnetization, $H_{eff}$ the effective magnetic field, $\alpha$ the damping factor, $\gamma$ the gyromagnetic ratio and s the unit vector parallel to M.

The damping factor $\alpha$ is generally assumed to be a scalar quantity [8, 9]. It is defined for small angle precession relaxation which is the case of FMR experiments. This parameter can be experimentally determined by FMR spectroscopy either from the angular variation of the linewidth or from the variation of the uniform mode linewidth $\Delta H_{pp}$ with the microwave frequency. In this second case the linewidth is given by:

$$\Delta H^{pp}(\omega) = \Delta H_{inhom} + \Delta H_{hom} = \Delta H_{inhom} + \frac{2}{\sqrt{3}}\cdot\frac{G}{\gamma^2\cdot M}\cdot\omega \qquad \text{eq.2}$$

With $\Delta H^{pp}$ the first derivative peak-to-peak linewidth of the uniform mode of Lorentzian lineshape, $\omega$ the angular microwave frequency and G the Gilbert damping factor from which the magnetization independent damping factor $\alpha$ can be deduced as $\alpha = G/\gamma M$. In eq. 2 it is assumed that the magnetization and the applied magnetic field are collinear which is fulfilled for high symmetry directions in GaMnAs such as [001], [110] and [100]. Otherwise a $1/\cos(\theta-\theta_H)$ term has to be added to eq.2 [8].

$\Delta H_{inhom}$ is the inhomogeneous, frequency independent linewidth; it can be further decomposed in three contributions, related to the crystalline imperfection of the film [10]:

$$\Delta H_{inhom} = \left|\frac{\delta H_r}{\partial\theta_H}\right|\cdot\Delta\theta_H + \left|\frac{\delta H_r}{\delta\phi_H}\right|\cdot\Delta\phi_H + \left|\frac{\delta H_r}{\delta H_{int}}\right|\cdot\Delta H_{int} \qquad \text{eq.3}$$

These three terms were introduced to take into account a slight mosaic structure of the metallic thin films defined by the polar angles ($\theta$, $\phi$) and their distributions ($\Delta\theta$, $\Delta\phi$) - expressed in the first two terms in eq.3- and a distribution of the internal anisotropy fields $H_{int}$

– the last term of eq.3. In the case of homoepitaxial III-V films obtained by MBE growth like GaMnAs on GaAs, films of high crystalline quality are obtained [LPN] and only the third component ($\Delta H_{int}$) is expected to play an important role.

Practically, the variation of the FMR linewidth with the microwave frequency can be measured with resonant cavity systems at different frequencies between 9GHz and 35GHz; the minimum requirement -used also in this work- is the use of two frequencies. We disposed in this work of 9GHz and 35GHz spectrometers. The linewidth is decomposed in a frequency independent inhomogeneous part and a linear frequency dependent homogeneously broadened part. For most materials the inhomogeneous fraction of the linewidth is strongly sample dependent and depends further on the interface quality and the presence of cap layers. It can be smaller but also much larger than the intrinsic linewidth. In $Ga_{0.95}Mn_{0.05}As$ single films total X-band linewidths between 100Oe and 1000Oe have been encountered. These observations indicate already the importance of inhomogeneous broadening. The homogeneous linewidth will depend on the intrinsic sample properties. This approach supposes that the inhomogeneous linewidth is frequency independent and the homogenous linewidth linear dependent on the frequency, two assumptions generally valid for high symmetry orientations of the applied field for which the magnetization is parallel to the magnetic field.

It should be underlined that in diluted magnetic semiconductor (DMS) materials like GaMnAs the damping parameter is not only determined by the sample composition $x_{Mn}$ [5]. It is expected to depend as well on (i) the magnetic compensation which will vary with the growth conditions, (ii) the (hole) carrier concentration responsible for the ferromagnetic Mn-Mn interaction which is influenced by the presence of native donor defects like arsenic antisite defects or Mn interstitial ions [11] and (iii) the valence bandstructure, sensitive to the strain in the film. Due to the high out-of–plane and in-plane anisotropy of the magnetic parameters [12] which further vary with the applied field and the temperature a rather complex situation with an anisotropic and temperature dependent damping factor can be expected in GaMnAs.

Whereas the FMR Gilbert damping factor has been determined for many metallic ferromagnetic thin films [8] only three experimental FMR studies have been published for GaMnAs thin films up to now [5-7]. In ref.[5,7] low doped GaMnAs layers with a critical temperature of 80K which do not correspond to the high quality, standard layers available today were studied. In the other work [6] higher doped layers were investigated but the experiments were limited to a single microwave frequency (9GHz) and thus no frequency dependence could be studied. In this work we present the results of FMR studies at 9GHz and 35 GHz on two high quality GaMnAs layers with optimum critical temperatures: one is a compressively strained layer grown on a GaAs buffer layer and the other a tensile strained layer grown on a (Ga,In)As buffer layer. Due to the opposite sign of the strains the easy axis of magnetization is in-plane [100] in the first case and out-of-plane [001] in the second. The GaMnAs layers have been annealed ex-situ after their growth in order to reduce the electrical and magnetic compensation, to homogenize the layers and to increase the Curie temperature to ≈ 130K. Such annealings have become a standard procedure for improving the magnetic properties of low temperature molecular beam epitaxy (LTMBE) grown GaMnAs films. Indeed, the low growth temperature required to incorporate the high Mn concentration without the formation of precipitates gives rise to native defect the concentration of which can be strongly reduced by the annealing.

## Experimental details

A first sample consisting of a $Ga_{0.93}Mn_{0.07}As$ layer of 50nm thickness has been grown at 250° C by low temperature molecular beam epitaxy on a semi-insulating (100) oriented GaAs substrate. A thin GaAs buffer layer has been grown before the deposition of the magnetic layer. The second sample, a 50 nm thick $Ga_{0.93}Mn_{0.07}As$ layer have been grown under very similar conditions on a partially relaxed (100) $Ga_{0.902}In_{0.098}As$ buffer layer; for more details see ref. [13]. After the growth the structure was thermally annealed at 250° C for 1h under air or nitrogen gas flow. The Curie temperatures were 157K and 130K respectively. Based on conductivity measurements the hole concentration is estimated in the $10^{20}cm^{-3}$ range.

The FMR measurements were performed with Bruker X-band and Q-band spectrometers under standard conditions: mW microwave power and 100 KHz field modulation. The samples were measured at temperatures between 4K and 170K. The angular variation of the FMR spectra was measured in the two rotation planes (110) and (001). The peak-to peak linewidth of the first derivative spectra were obtained from a lineshape simulation. The value of the static magnetization M(T) had been determined by a commercial superconducting quantum interference device (SQUID) magnetometer. A typical hysteresis curve is shown in the inset of fig.8.

## Experimental results:

The saturation magnetizations of the two layers and the magneto crystalline anisotropy constants which had been previously determined by SQUID and FMR measurements respectively are given in table I. The anisotropy constants had been determined in the whole temperature range but for clarity only its values at T=55K and T=80K are given in table I. We see that the dominant anisotropy constant $K_{2\perp}$ are of different sign with -55000 erg/cm$^3$ to +91070 erg /cm$^3$ and that the other three constants have equally opposite signs in the two types of layers. The easy axes of magnetization are the in-plane [100] and the out-of-plane [001] direction respectively. However the absolute values of the total effective perpendicular anisotropy constant $Ku=K_{2\perp}+K_{4\perp}$ are less different for the two samples: -46517erg/cm$^3$ and +57020erg/cm$^3$ respectively. More detailed information on the measurements of these micromagnetic parameters will be published elsewhere.

For the GaMnAs/GaAs layers the peak-to-peak linewidth of the first derivative uniform mode spectra has been strongly reduced by the thermal annealing; in the non annealed sample the X-band linewidth was highly anisotropic with values between 150Oe and 500Oe at T=4K. After annealing it is reduced to an quasi isotropic average value of 70Oe at X-band. Quite differently, for the GaMnAs/GaInAs system the annealing process decreases the linewidth of the GaMnAs layers only marginally. Although full angular dependencies have been measured by FMR we will analyze only the linewidth of the four high symmetry field orientations H//[001], H//[100], H//[1-10], H//[110] corresponding to the hard and easy axes of magnetization. As will be shown below, in spite of rather similar high critical temperatures (157K/130K) the linewidth are drastically different for the two cases.

**1. GaMnAs on GaAs**

In fig. 1a and 1b we show typical low temperature FMR spectra at X-band and Q-band frequencies for the hard [001] /intermediate [100] axis orientation of the applied magnetic field. The spectra are characterized by excellent signal to noise ratios and well defined lineshapes. We see that at both frequencies the lineshape is close to a Lorentzian. In addition to the main mode one low intensity spin wave resonance is observed at both frequencies at lower fields (not shown).

The linewidth at X-band (fig.2) is of the order of 50Oe to 75Oe with a weak orientation and temperature dependence. Above T>130K, close to the critical temperature, the linewidth increase strongly. At Q-band we observe a systematic increase by a factor of two of the total linewidth (fig.3) with an increased temperature and orientation dependence. As generally observed in GaMnAs, the easy axis orientation gives rise to the lowest linewidth. At Q-band the lineshape is perfectly Lorentzian (fig.1b). These linewidth are among the smallest ever reported for GaMnAs thin films, which reflects the high crystalline and magnetic quality of the film.

To determine the damping factor α we have plotted the frequency dependence of the linewidth for the different orientations and at various temperatures. An example is given in fig. 4 for T=80K; this allows us to determine the inhomogeneous linewidth obtained from a linear extrapolation to zero frequency and the damping factor from the slope. The inhomogeneous linewidth at T=80K is of the order of 30 Oe, i.e. 50% of the total linewidth at X-band. This shows that the approximation $\Delta H_{inhom} << \Delta H_{homo}$ which had been previously used [5] to deduce the damping factor from a single (X-band) frequency measurement is not fulfilled here.

The temperature dependence of the inhomogeneous linewidth is shown in fig.5. Similar trends as for the total linewidth in the non annealed films are observed: the linewidth is high at the lowest temperatures, decreases with increasing temperatures up to 120K and increases again close to $T_C$.

From the slope of the linewidth variation with microwave frequency we obtain the damping factor α (fig.6). Its high temperature value is of the order of 0.010 but we observe a systematic, linear variation with the temperature and a factor two difference between the easy axis orientation [100] and the hard axis orientation [001].

**2. GaMnAs on GaInAs**

Similar measurements have been performed on the annealed tensile strained layer. In tensile strained GaMnAs films the easy axis of magnetization ([001]) coincides with the strong uniaxial second order anisotropy direction. For that reason no FMR resonance can be observed at temperatures below T=80K for the easy axis orientation H// [001] at X-band. For the other three orientations the resonances can be observed at X-band in the whole temperature range 4K to $T_C$. Due to the strong temperature dependence of the anisotropy constants and the parallel decrease of the internal anisotropy fields the easy axis resonance becomes observable at X-band for temperatures above 80K. In the films on GaInAs much higher linewidth are encountered than in films on GaAs, the values are up to ten times higher indicating a strong inhomogeneity in this film. A second low field resonance is systematically observed at X-band and Q-band; it is equally attributed to a spin wave resonance.

Figures 7a and 7b show typical FMR spectra at X- and Q-band respectively. At both frequencies the lineshape can no longer be simulated by a Lorentzian but has changed into a Gaussian lineshape.

Contrary to the first case of GaMnAs/GaAs the X-band linewidth varies monotonously in the whole temperature region (fig.8). We observe a linewidth of ~600Oe at T=4K, which decreases only slowly with temperature; the linewidth becomes minimal in the 100 K to 140K range. The Curie temperature "seen" by the FMR spectroscopy is slightly higher as compared to the one measured by SQUID due to the presence of the applied magnetic field.

At low temperature the Q-band linewidth vary strongly with the orientation of the applied field with values between 500Oe and 700Oe. The lowest value is observed for the easy axis orientation. They decrease as at X-band only slowly with increasing temperature and increase once again when approaching the Curie temperature. At Q-band the easy axis FMR spectrum, which is also accompanied by a strong spin wave spectrum at lower fields, is observable in the whole temperature range.

For this sample we observe especially at Q-band a systematic difference between the cubic axes [100], [001] linewidth and the one for the in-plane [110] and [1-10] field orientations (fig.8). The most surprising observation is that for temperatures below T<100K the linewidth for H//[100] and H//[110] are comparable at X-band and Q-band and thus an analysis in the simple model discussed above is not possible. We attribute this to much higher crystallographic/magnetic inhomogeneities, which mask the homogenous linewidth. The origin of the strong inhomogeneity is still unclear. The only orientation for which in the whole temperature range a systematic increase in the linewidth between X-and Q-band is observed is the H//[1-10] orientation. We have thus analyzed this variation (fig.10) according to eq.1.

In spite of important differences in the linewidth the slope varies only weakly which indicates that the inhomogeneous linewidth is very temperature dependent and decreases monotonously with increasing temperature from 570Oe to 350Oe.

In the high temperature range (T≥100K) the easy axis orientation could also be analyzed (fig.11). The inhomogeneous linewidth are lower than for the hard axis orientation at the same temperatures and are in the 300Oe range (fig.12). The homogenous linewidth at 9Ghz is in the 50Oe range which is close to the values determined in the first case of GaMnAs/GaAs.

From the slope (fig.13) we obtain the damping factor which for the hard axis orientation is $\alpha$=0.010 in the whole temperature range. This value is comparable to the one measured for the GaMnAs/GaAs film for H//[110]. The damping factor for the easy axis orientation is lower but increases close to $T_C$ as in the previous case.

**Discussion:**

An estimation of the FMR intrinsic damping factor in a ferromagnetic GaMnAs thin film has been made within a model of localized Mn spins coupled by p-d kinetic exchange with the itinerant-spin of holes treated by the 6-band Kohn-Luttinger Hamiltonian [5]. Note, that these authors take for the effective kinetic exchange field the value in the mean-field approximation, i.e. $H_{eff}$=JN<S>, so that their calculation are made within the random phase approximation (RPA). RPA calculations of $\alpha$ have been made by Heinrich et al. [14] and have recently been used by Tserkovnyak et al.[15] for numerical applications to the case of $Ga_{0.95}Mn_{0.05}As$. Both models however, are phenomenological and include an adjustable parameter: the quasiparticle lifetime $\Gamma$ for the holes in [5] and the spin-flip relaxation $T_2$ in [15]. These models do not take into account neither multi-magnon scattering nor any damping beyond the RPA. It has been argued elsewhere [16], that in diluted magnetic semiconductors such affects are only important at high temperature (i.e. at T>Tc). In particular, the increase of $\alpha$ in the vicinity of Tc may be attributed to such effects that are beyond the scope of the models of references [5] and [14]. At low temperatures T<<Tc however, where the corrections to the RPA are expected to be negligible, the models of [5, 14,15] provide us with a numerical value of $\alpha$ in agreement with our experiments if we introduce reasonable values of these parameters. For GaMnAs films with metallic conductivity, Mn concentrations of x=0.05 and hole concentrations of 0.5 $nm^{-3}$ ($5x10^{20}cm^{-3}$) Sinova et al [5] predict an isotropic low temperature damping factor $\alpha$ between 0.02 and 0.03 depending on the quasiparticle life time broadening. Tserkovnyak et al [15] found a similar value of $\alpha \approx 0.01$ for the isotropic damping factor for a typical GaMnAs film with 5% Mn doping and full hole polarization.

Both predicted values are of the same order of magnitude as the experimental values determined in this study. Our results for GaMnAs/GaAs show further that the damping factor is not isotropic as generally assumed but is anisotropic with a lowest value for the in-plane easy axis orientations of the applied magnetic field H//[100], H//[1-10] and an increase of up to a factor of two for the hardest axis orientation H// [001]. Intrinsic anisotropic damping is related to the fact that the free energy density depends on the orientation of the magnetization which in the case of GaMnAs is related to the anisotropy of the p-hole Fermi surface. We have shown (table I) that the anisotropy of the magnetocristalline constants and the related fields are important in these strained layers and it is thus not surprising to find also anisotropy of the damping factor. For further discussions on this subject see reference [9]. The system might also contain extrinsic anisotropies related to the presence of lattice defects. Their influence can be deduced from the value and anisotropy of the inhomogeneous linewidth. In the case of the compressively strained layers (GaMnAs/GaAs) we see that their value is small and rather isotropic quite differently from the tensile strained film. It is in the first case that our measurements show a factor of two anisotropy of the Gilbert damping factors. A further indication for the intrinsic character of the anisotropy is the fact that the damping factor for the perpendicular orientation has the highest value. In this case any contribution from two magnon scattering will be minimized. Anyway, such contributions are generally only important at low frequency measurements in the 2-6GHz range but even there they were found to be negligible [7].

Additional material related parameters are expected to further influence the damping factor. As the spin flip relaxation times will depend on the sample properties and in particular the presence of scattering centers we will not expect to find a unique damping factor even for GaMnAs/GaAs samples with the same Mn composition x. More likely, different damping factors are expected to be found in real films and their values might be used to assess the film quality. In this sense the GaMnAs/GaAs film studied here is of course "better" than the one on GaInAs in line with the strong difference in the sample inhomogeneities.

The inhomogeneous linewidth originates from spatial inhomogeneities in the local magnetic anisotropy fields and inhomogeneities in the local exchange interactions. Given the particular growth conditions of these films, low temperature molecular beam epitaxy, inhomogeneities can not be expected to be negligible in these materials. If we had analyzed our X-band results of the GaMnAs/GaAs film in the spirit of ref. [5], i.e. assuming a negligible inhomogeneous linewidth - ($\Delta H_{inhom}=0$) -, we would have obtained artificially increased damping factors. A further contribution might be expected from the intrinsic disorder in these films: as GaMnAs is a diluted magnetic semiconductor with random distribution of the Mn ions, this disorder will even for crystallographically perfect crystals give some importance to this term.

In the previous studies of the FMR damping factor in GaMnAs/GaAs single films higher values have been reported. Matsuda et al [6] found damping factors between 0.02 and 0.06 in the T=10K to T=20K temperature range. They observed the same tendencies as in this work concerning the anisotropy and temperature dependence of $\alpha$: the lowest damping factor is seen for the easy axis orientation and its values increases with increasing temperatures. The films of their study were however significantly different: (i) the Mn doping concentration was lower, x=0.03 and thus the hole concentration was equally lower and (ii) the critical temperature of the annealed film was only T=80K. The anisotropy in the inhomogeneous linewidth at T=20K was equally much higher, varying between 30Oe for the easy axis to 250Oe for the hardest axis [110]. In a second study Sinova et al [5] have measured an annealed GaMnAs/GaAs sample with a similar composition (x= 0.08) and critical temperature ($T_C$=130K) as the one studied here. They deduced a damping factor of the order of $\alpha$=0.025 with only a slight temperature dependence between 4K and 80K and an increase close to $T_C$. However, these measurements were done at one (X-band) microwave frequency only and the numerical value of $\alpha$ was obtained by assuming a negligible inhomogeneous linewidth. As explained above, the value of $\alpha$ can be expected to be overestimated in this case.In the photovoltage measurements of ref.7 the damping factor of a low (x=0.02) doped GaMnAs layer has been determined with microwave frequencies from 2 to 19.6GHz but for one field orientation H//[001] (hard axis) and one temperature (T=9K) only. Interestingly, their measurements show a linear behavior even in the low frequency range down to 4GHz which demonstrates the negligible contribution from two magnon scattering in this case.

The intrinsic damping factor $\alpha$ plays also an important role in the critical currents required to switch the magnetization in FM/NM/FM trilayers [5]. However, in trilayer structures interface and spin pumping effects will add to the intrinsic damping factor of the ferromagnetic material and give rise to an increased effective damping factor. Sinova et al [5] have estimated the critical current for realistic GaMnAs layers: based on a value of $\alpha$= 0.02, they estimated the critical current density to $J_C=10^5 A/cm^2$. It should be noted that the damping factor involved in the domain wall motion [17, 18] is by definition different form the FMR damping factor. Both are however linearly related with $\alpha_{FMR}<\alpha_{DW}$ [17, 18]. First observations of current induced magnetization switching in $Ga_{0.956}Mn_{0.044}As/GaAs/Ga_{0.967}Mn_{0.033}As$ trilayers confirm these theoretical predictions [19]. These authors observed a critical current density of $\approx 10^5 A/cm^2$ which would have been predicted from Slonczewski's formula [20] for a domain wall damping factor of $\alpha_{DW}$=0.002.

**Conclusion:**

We have determined the intrinsic FMR Gilbert damping factor for annealed $Ga_{0.93}Mn_{0.07}As$ thin films with high critical temperatures. To evaluate the influence of the strain the two prototype cases of compressive and tensile strained layers were studied. In both cases we find an average damping factor of the order of 0.01. We thus see that the sign of the strain does not seem to influence the damping factor strongly. The homogeneity of the films as judged from the inhomogeneous linewidth is much higher in the case of GaAs substrates than for GaInAs substrates. This must be attributed to the high dislocation density in the GaInAs layer [13]. In the case of the GaMnAs/GaAs layers, where the small linewidth allows a finer analysis of the data, we observe an anisotropy of the damping factor, which has the lowest value for the easy axis orientation. This value of $\alpha_{[1-10]}=0.003$ is an order of magnitude lower than the previous reported values. The few experimental results available seem to indicate that the damping factor decreases with increasing Mn concentration. This corresponds well to the theoretical predictions by Sinova et al [5] for the case of small quasi-particle lifetime broadening. It will be interesting to test this behavior further in more highly doped layers ($x\approx0.15$), which become now available.

**Acknowledgment:** We thank Alain Mauger of the IMPMC laboratory of the University Paris 6 and Bret Heinrich from the Simon Frazer University in Vancouver for many helpful discussions.

## Figure Captions:

Figure 1a: X-band FMR spectrum of the GaMnAs/GaAs film taken at T=20K and for H// [001]; the peak-to-peak linewidth is 60Oe. The experimental spectrum is shown by circles and the Lorentzian lineshape simulation by a line.

Figure 1b: Q-band FMR spectrum of the GaMnAs/GaInAs film taken at T=80K and for H// [100]; the peak to peak linewidth is 120Oe. The experimental spectrum is shown by circles and the Lorentzian lineshape simulation by a line

Figure 2 (color on line): X-band peak-to-peak line widths for the GaMnAs/GaAs film for the four main field orientations: H//[001] black squares, H//[1-10] olive lower triangles, H//[110] red circles, H//[100] blue upper triangles.

Figure 3 color on line): Q-band peak-to-peak line widths for the GaMnAs/GaAs film for the four main field orientations: H//[001] black squares, H//[1-10] olive lower triangles, H//[110] red circles, H//[100] blue upper triangles.

Figure 4 (color on line): Peak-to-peak linewidth at 9GHz and 35GHz for the GaMnAs/GaAs film; T=80K and H//[001] black squares, H//[1-10] olive lower triangles, H//[110] red circles, H//[100] blue upper triangles

Figure 5 (color on line): GaMnAs/GaAs inhomogeneous linewidth as a function of temperature for four orientations of the applied field: H//[001] black squares, H//[1-10] olive lower triangles, H//[110] red circles, H//[100] blue upper triangles

Figure 6 (color on line): damping factor $\alpha$ as a function of temperature and magnetic field orientation; H//[001] black squares, H//[1-10] olive lower triangles, H//[110] blue upper triangles, H//[100] red circles; the maximum error in the determination of the linewidth is estimated to 10G which corresponds to an error in $\alpha$ of 0.001

Figure 7a: X-band FMR spectra for the GaMnAs/GaInAs film at 25K and H// [110] (hard axis); the low field spin wave resonance (SW) is of high intensity in this case; circles: experimental points, line: simulation with Gaussian lineshape

Figure 7b: Q-band FMR spectra for the GaMnAs/GaInAs film at 25K and H// [110] (hard axis); the low field spin wave resonance (SW) is of high intensity in this case; circles: experimental points, line: simulation with Gaussian lineshape

Figure 8 (color on line): GaMnAs/GaInAs: X-band FMR linewidth as a function of temperature for the four orientations of the applied magnetic field: H//[001] black squares, H//[1-10] olive lower triangles, H//[110] red circles, H//[100] blue upper triangles; the easy axis FMR spectrum is not observable below T=100K; a typical hysteresis curve as measured by SQUID is shown in the inset.

Figure 9 (color on line): GaMnAs/GaInAs : Q-band FMR linewidth as a function of temperature for the four orientations of the applied magnetic field : H//[001] black squares, H//[1-10] olive lower triangles, H//[110] red circles, H//[100] blue upper triangles

Figure 10 (color on line): GaMnAs/GaInAs : FMR linewidth as a function of microwave frequency for H//[1-10] at different temperatures; T=10K (black squares), T=25K (red circles), T=55K(black spades), T=80K(blue stars), T=115K (red triangles)

Figure 11 (color on line): GaMnAs/GaInAs: FMR linewidth as a function of microwave frequency for H//[001] (easy axis) at different temperatures; T=100K (red circles), T=120K (blue spades),T=130K(black squares)

Figure 12: GaMnAs/GaInAs inhomogeneous linewidth as a function of temperature for two orientations of the applied field: H//[001] squares, H//[1-10] lower triangles,

Figure 13: GaMnAs/GaInAs damping factor $\alpha$ as a function of temperature and magnetic field orientation; H//[001] squares, H//[1-10] lower triangles.

# Tables

| $Ga_{0.93}Mn_{0.07}As/GaAs$ | $Ga_{0.93}Mn_{0.07}As/Ga_{0.902}In_{0.098}As$ |
|---|---|
| Ms(T=4K)= 47emu/cm$^3$ | Ms(T=4K)= 38 emu/cm$^3$ |
| $T_C$ =157K (SQUID) | $T_C$=130K (SQUID) |
| **Anisotropy constants (T=80K)** | **Anisotropy constants (T=55K)** |
| $K_{2\perp}$ = - 55000 erg/cm$^3$ | $K_{2\perp}$ = +91070 erg/cm$^3$ |
| $K_{2//}$ =2617 erg/cm$^3$ | $K_{2//}$ = -2464 erg/cm$^3$ |
| $K_{4\perp}$ = 8483 erg/cm$^3$ | $K_{4\perp}$ = -34050 erg/cm$^3$ |
| $K_{2//}$ =2590 erg/cm$^3$ | $K_{2//}$ = -1873erg/cm$^3$ |
| **Easy axis of magnetization** | **Easy axis of magnetization** |
| [100] 4K<T<$T_C$ | [001] 4K<T< $T_C$ |

Table I

# Table caption

Table I: Micromagnetic parameters of the two samples studied in this work: saturation magnetization $M_s$ at T=4K, critical temperature $T_C$, magneto crystalline anisotropy constants of second and fourth order $K_{2\perp}$ , $K_{2//}$, $K_{4\perp}$, $K_{4//}$ at T=55K and T=80K respectively and the orientation of the easy axis for magnetization.

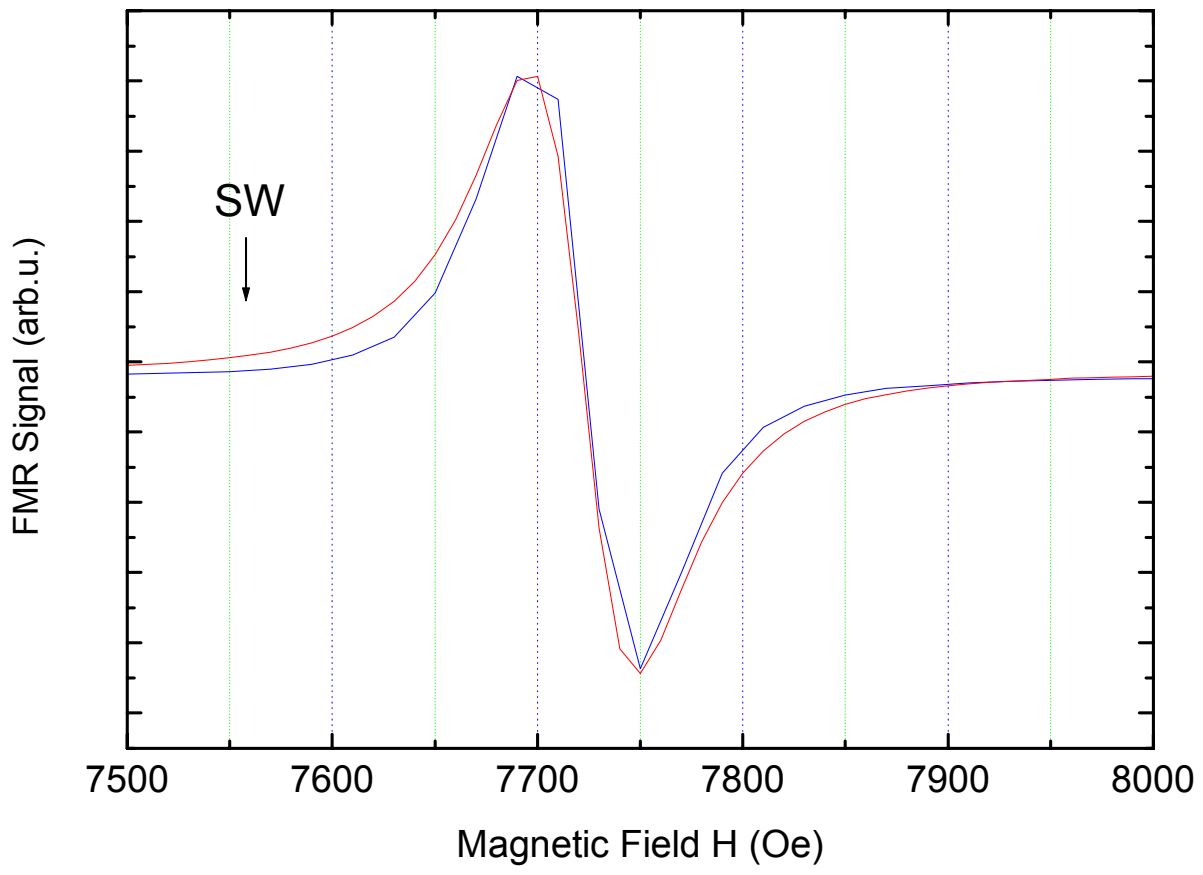

Figure 1a

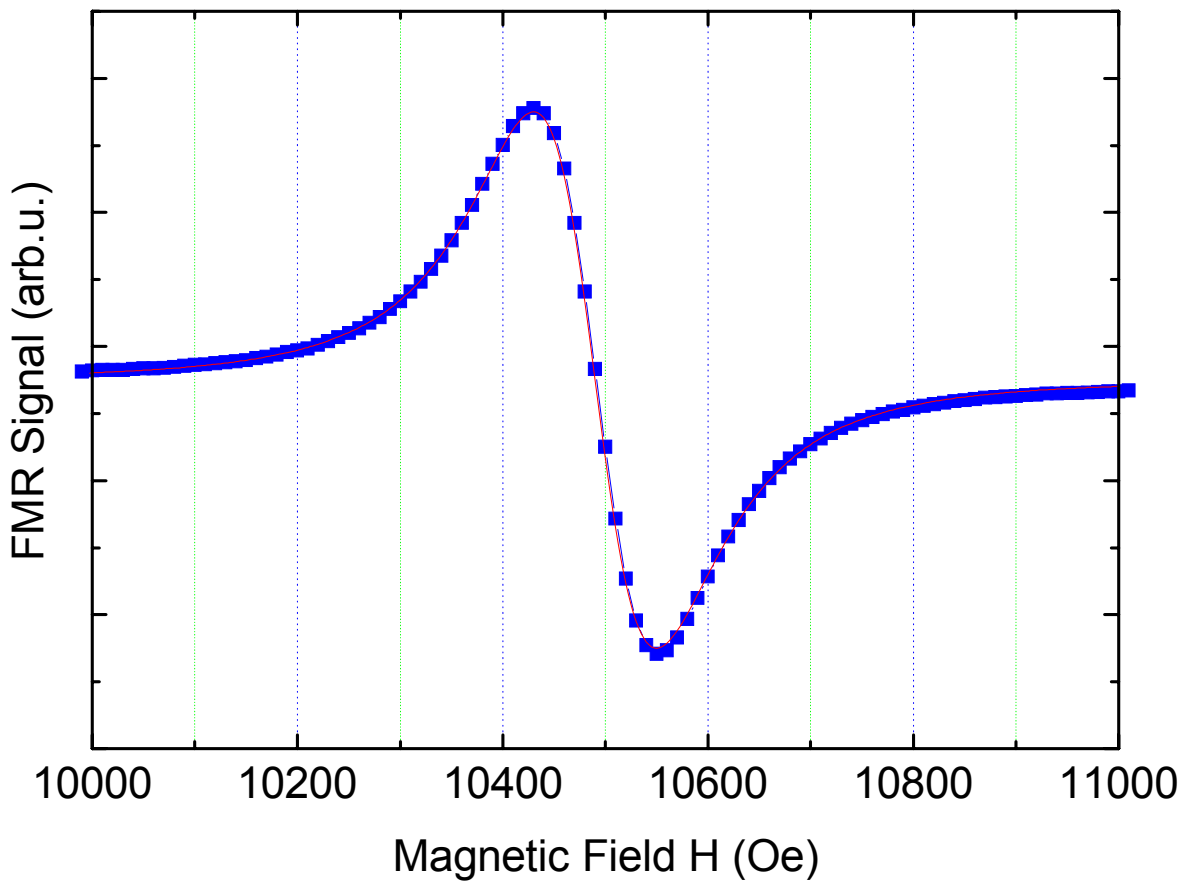

Figure 1b

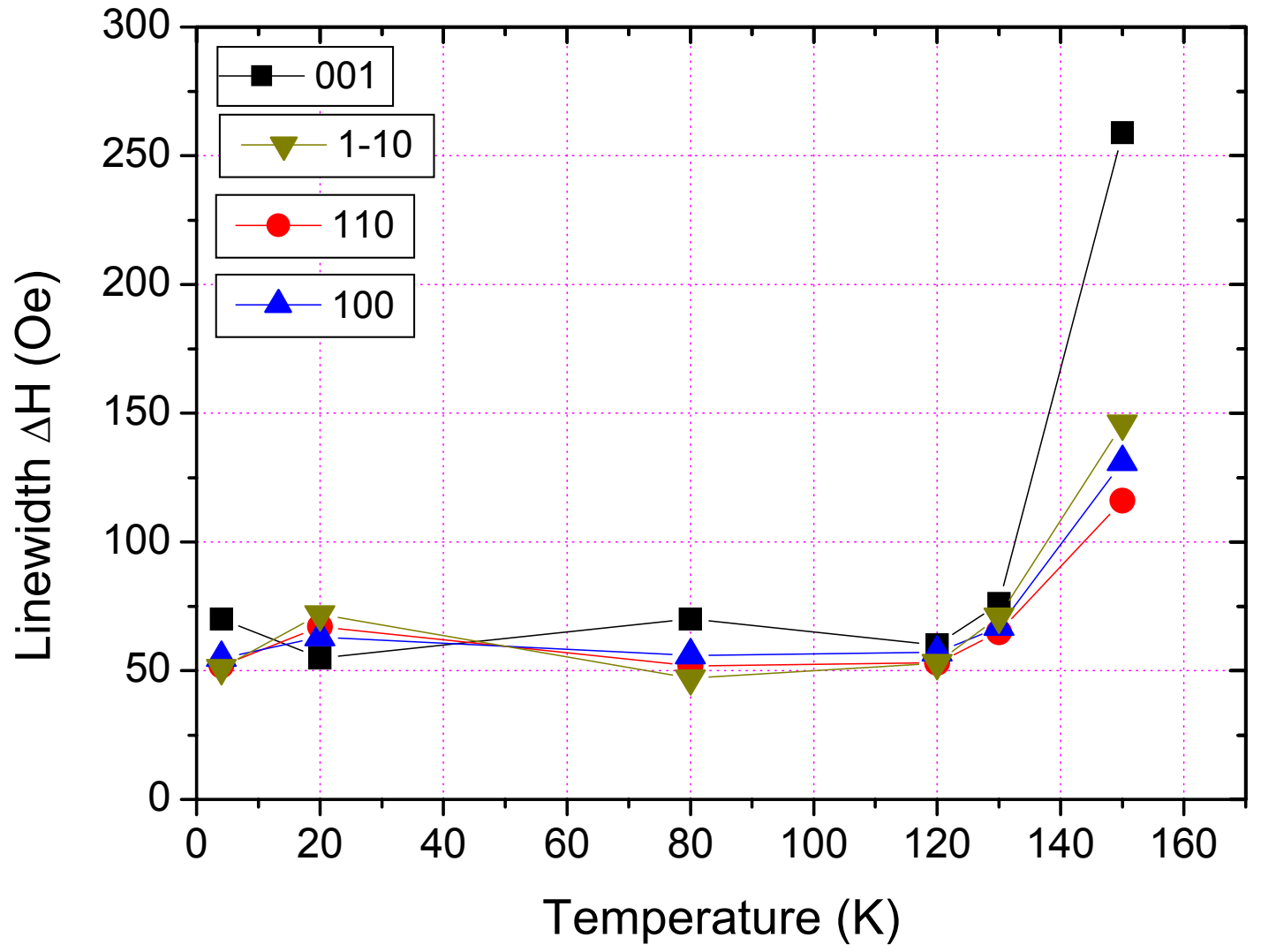

Figure 2

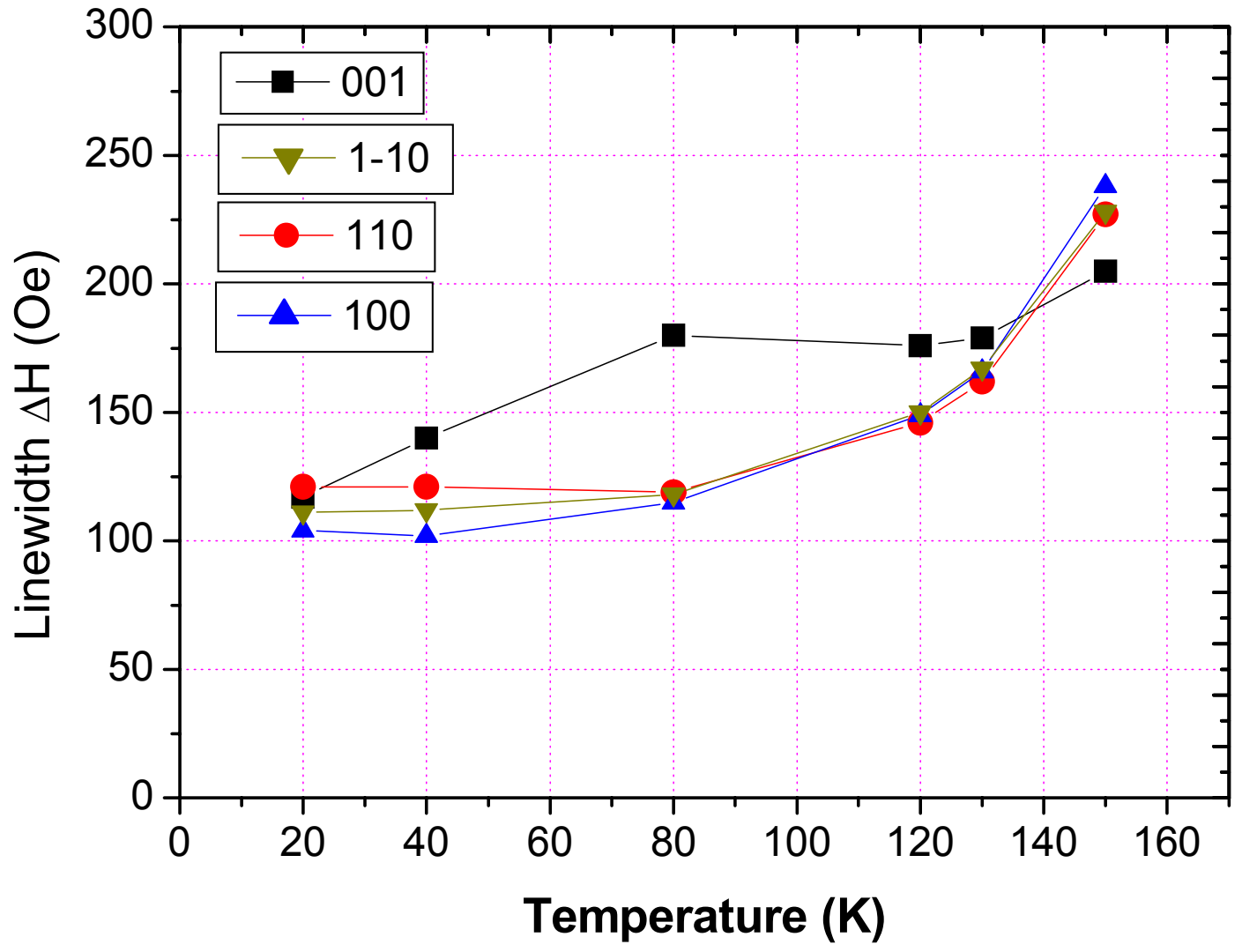

Figure 3

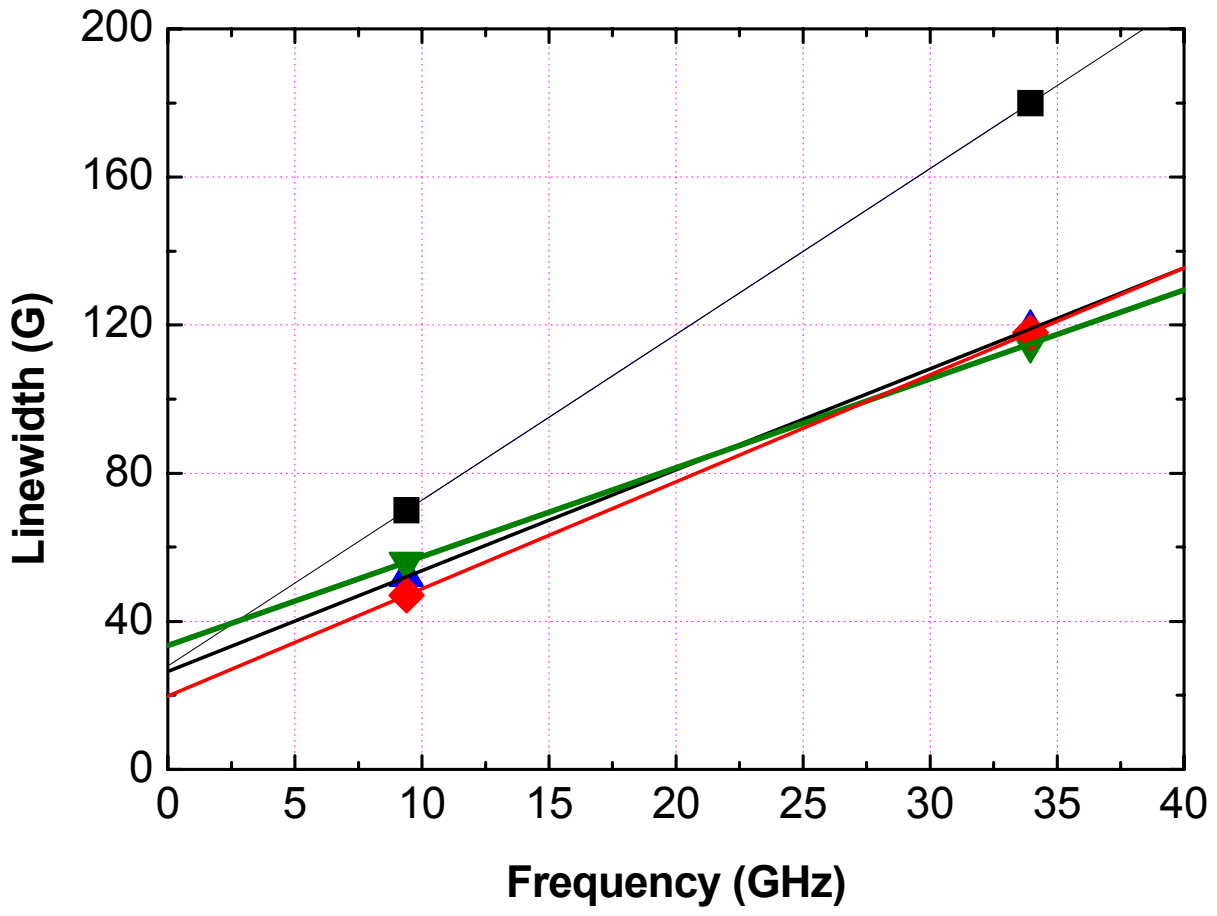

Figure 4

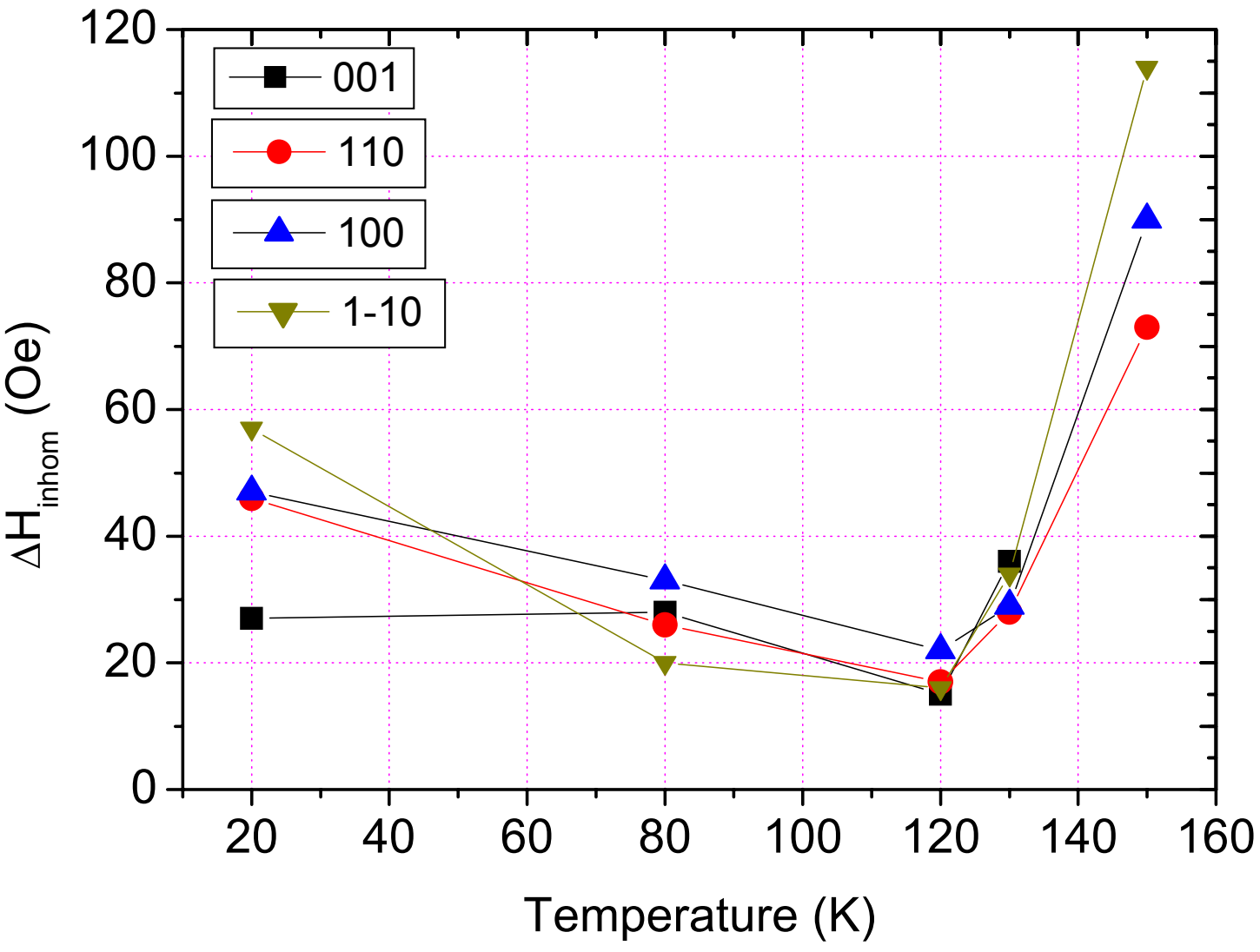

Figure 5

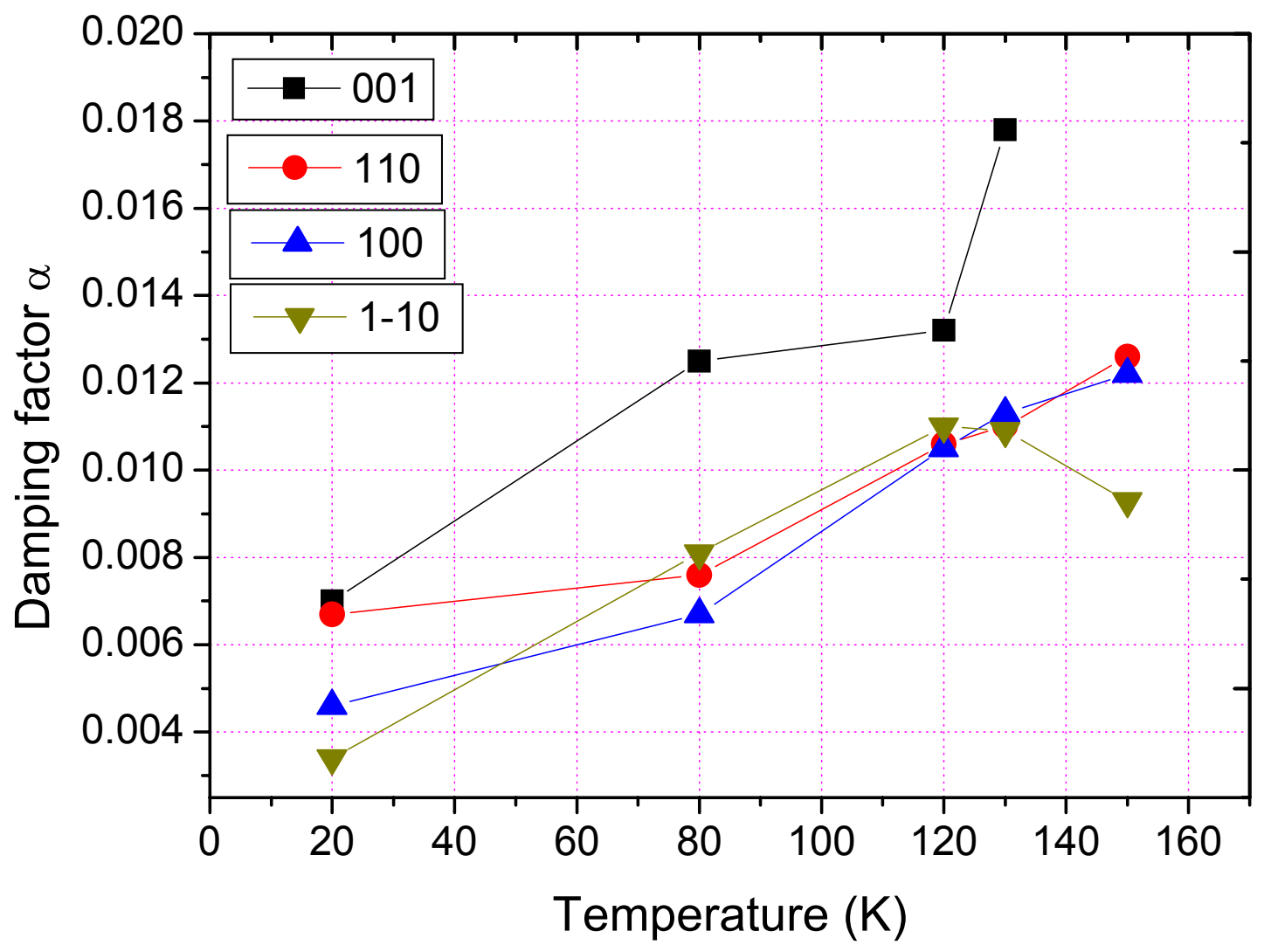

Figure 6

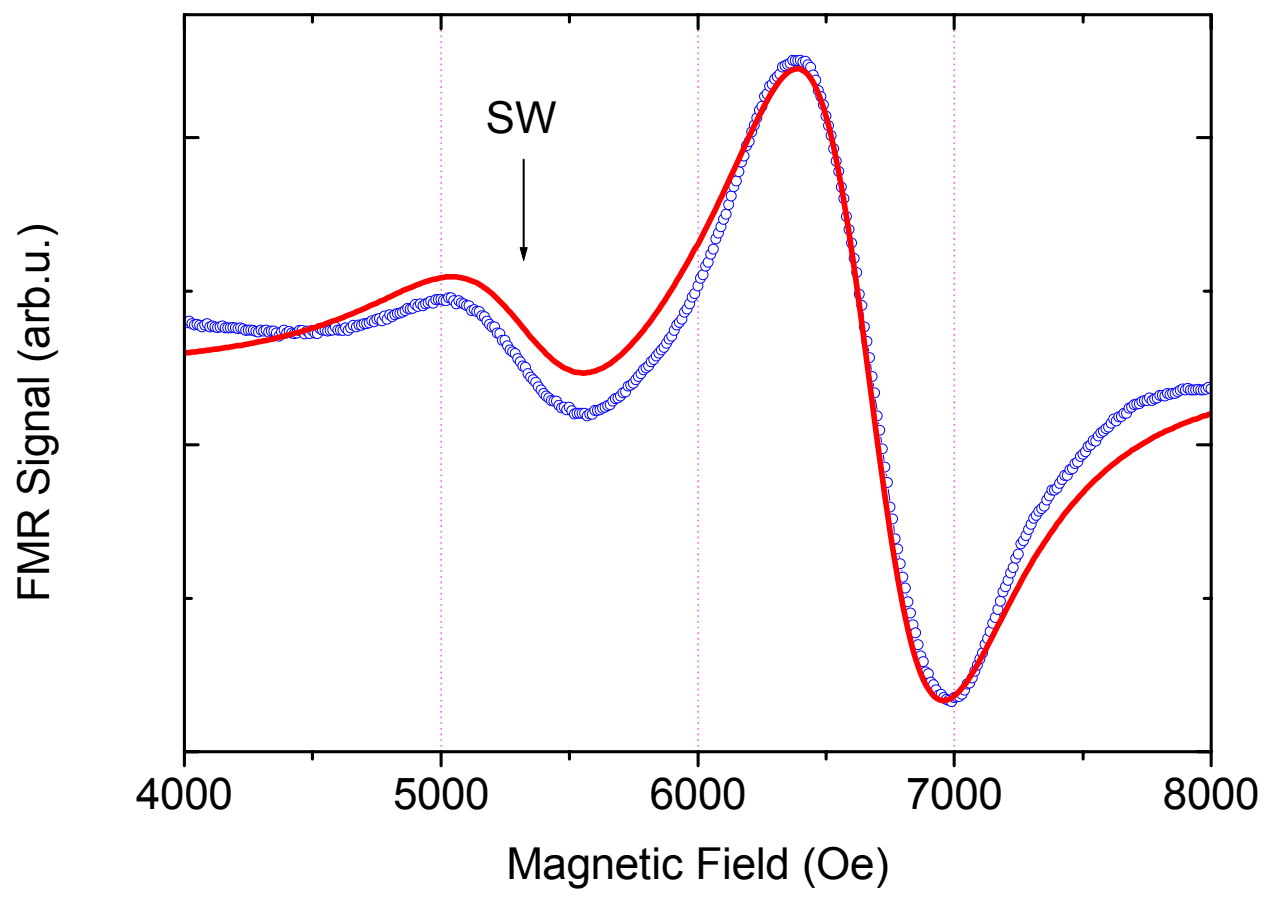

Figure 7a

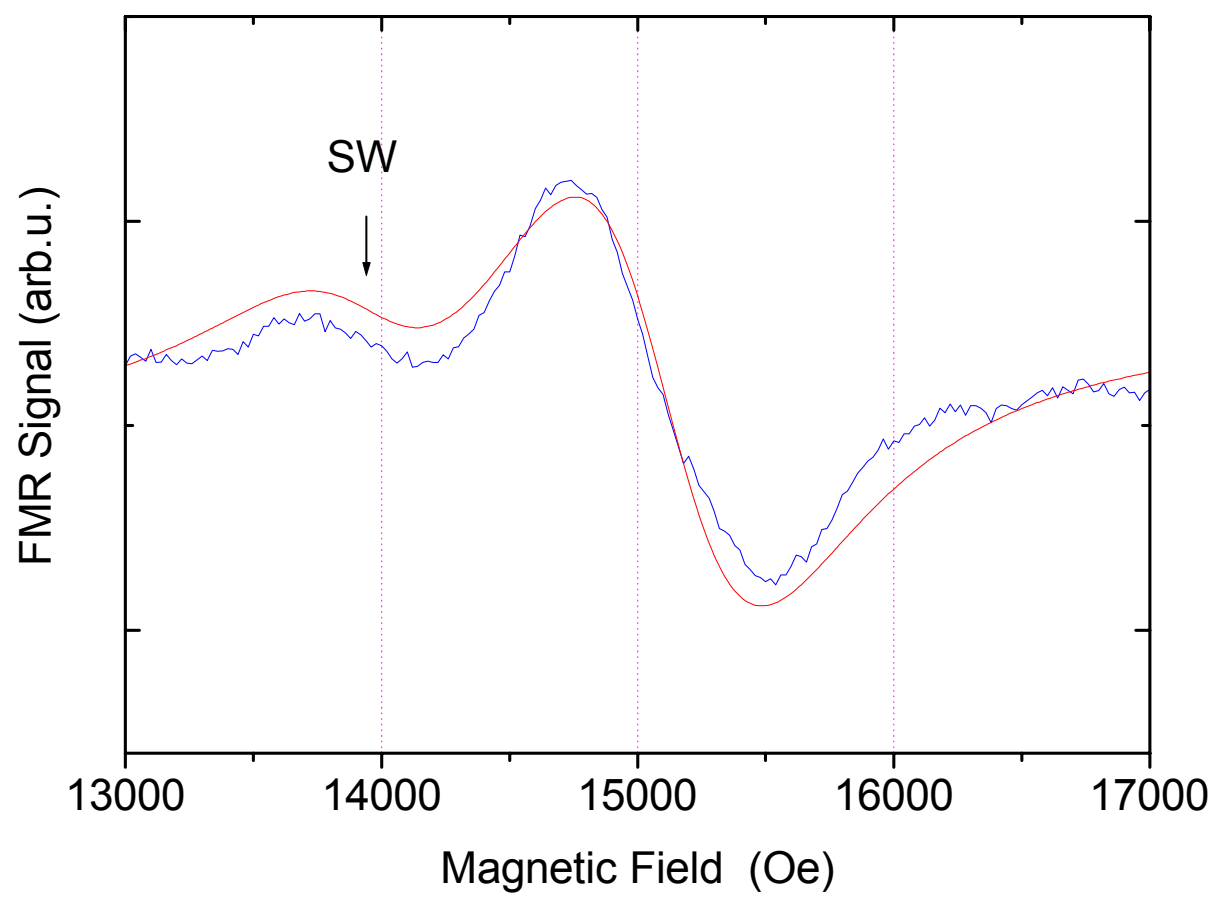

Figure 7b

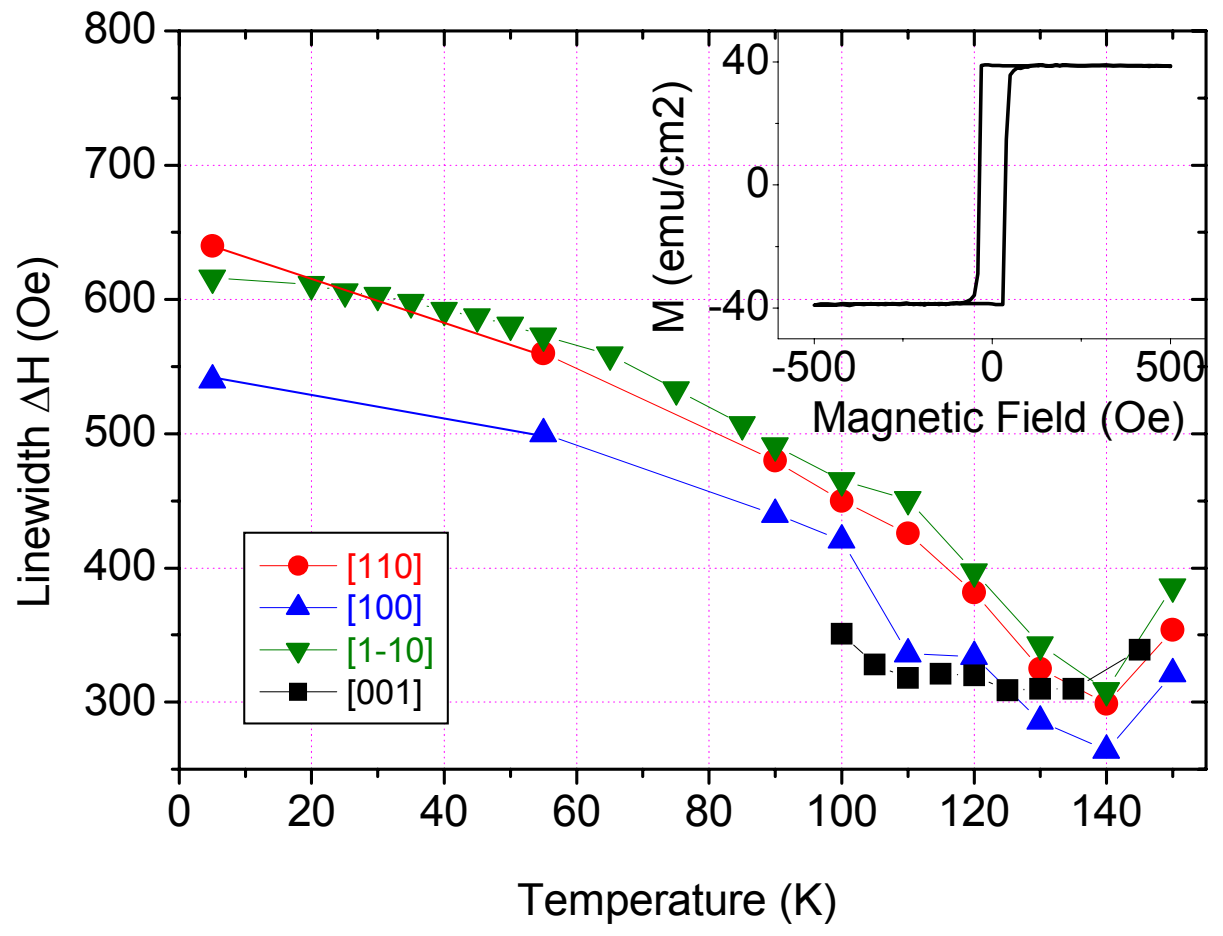

Figure 8

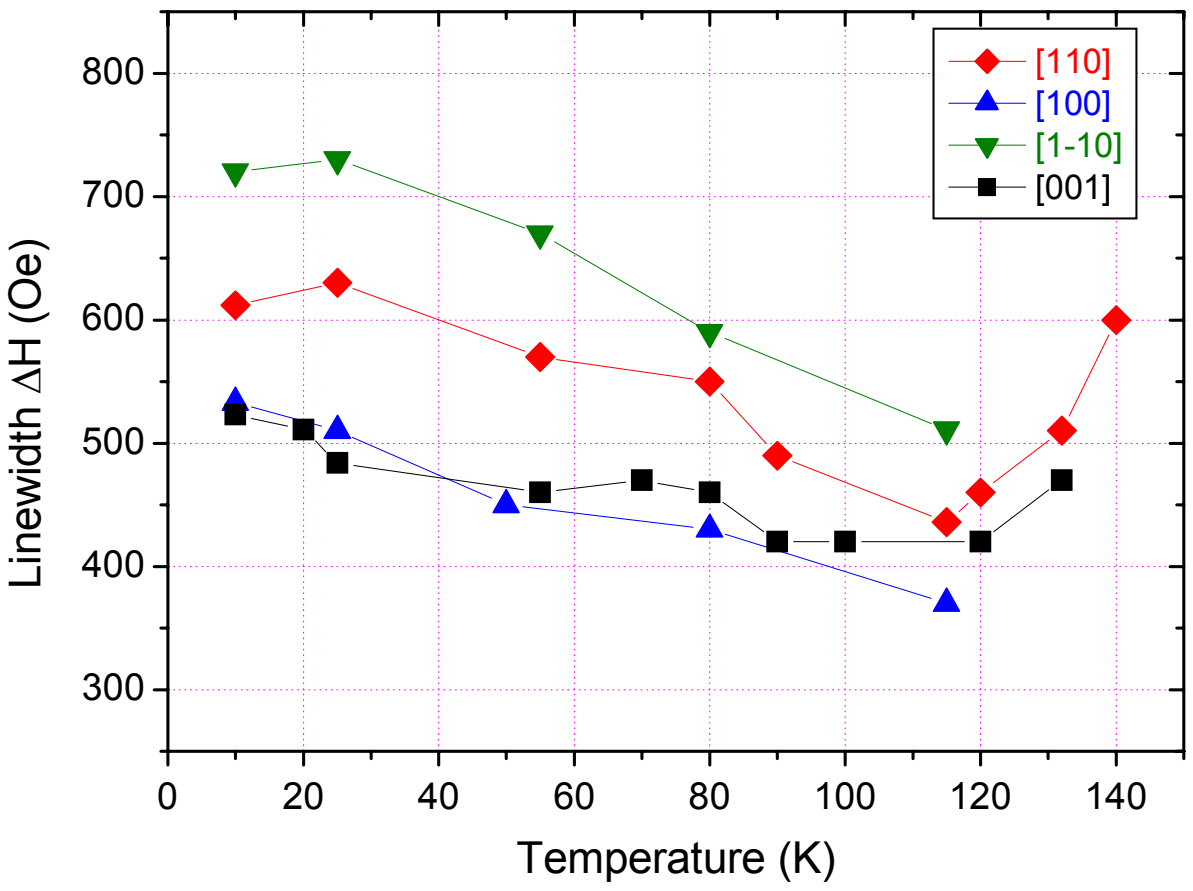

Figure 9

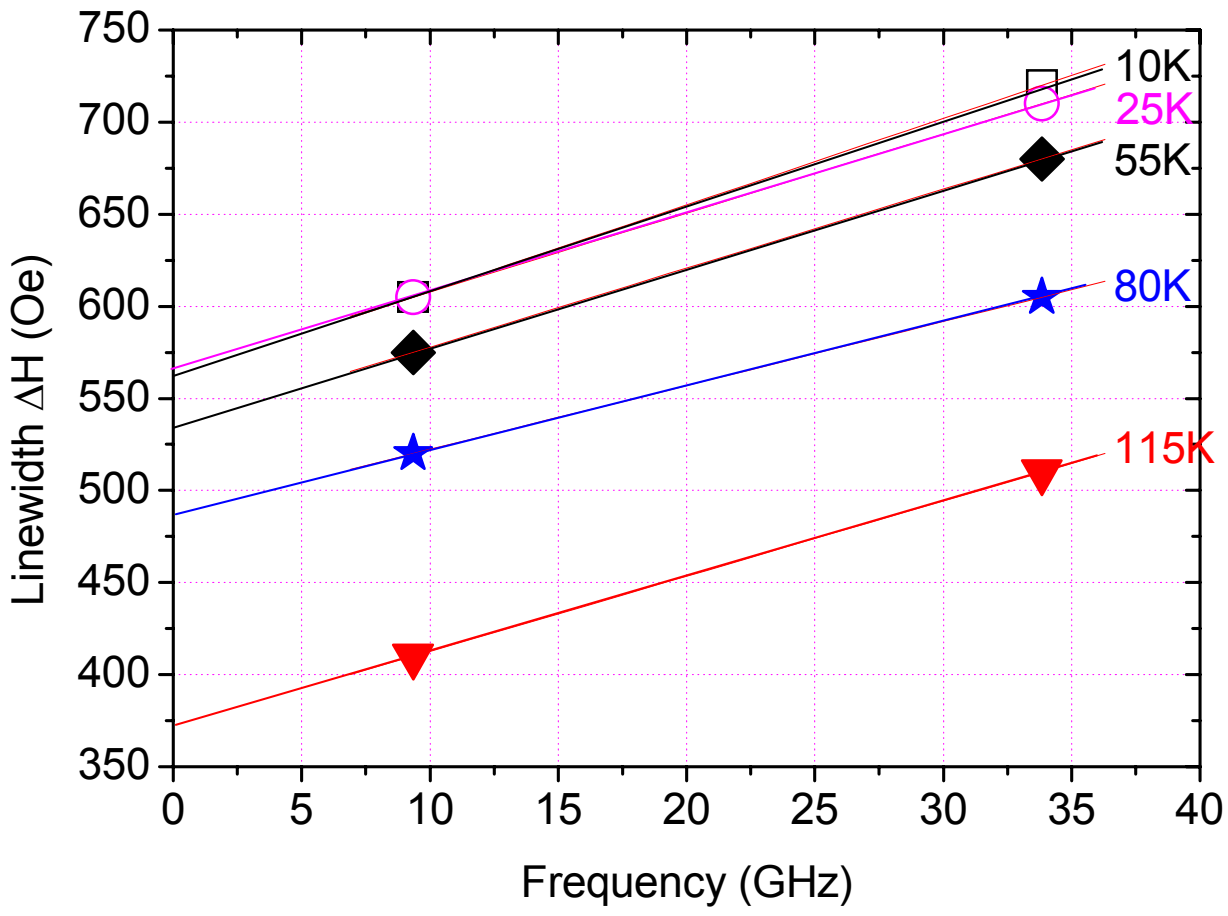

Figure 10

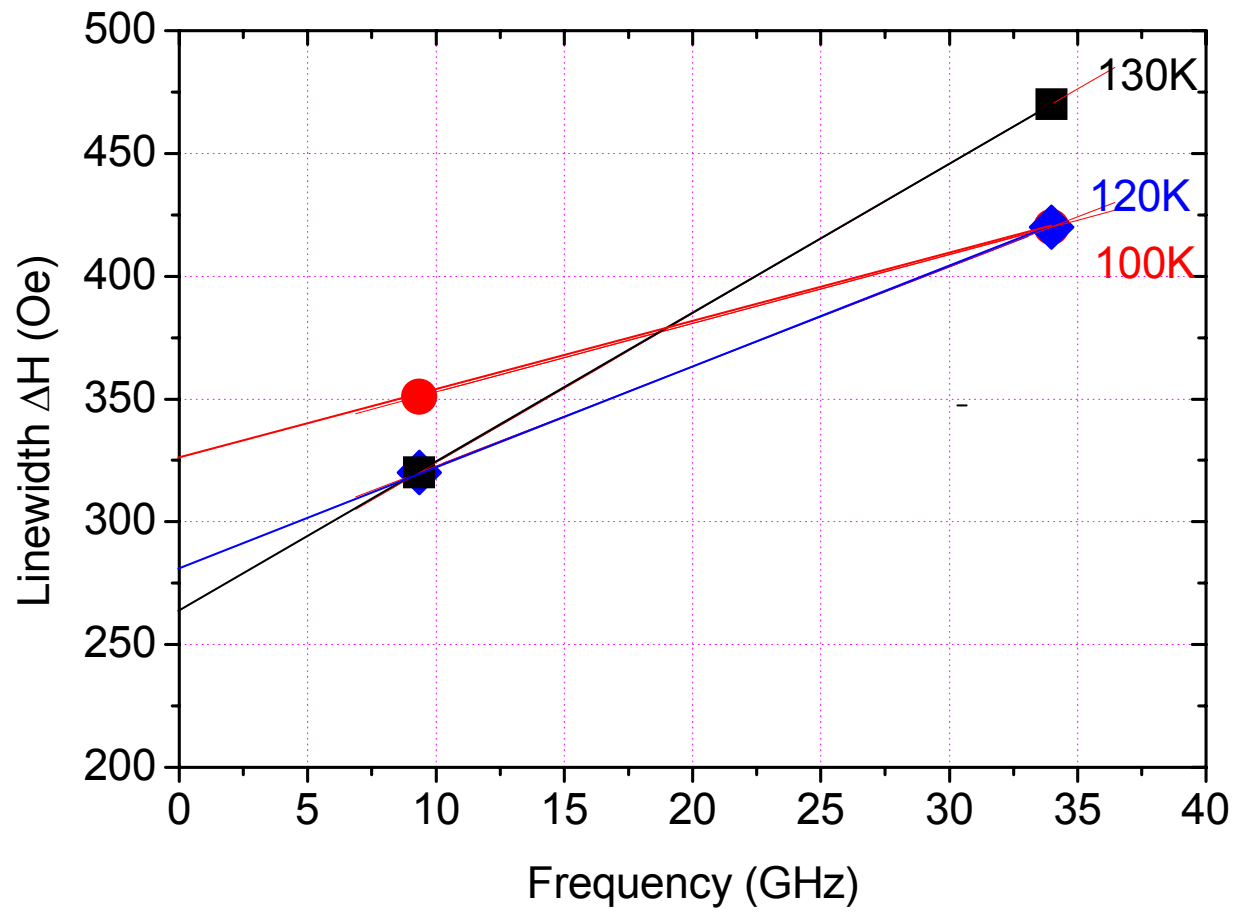

Figure 11

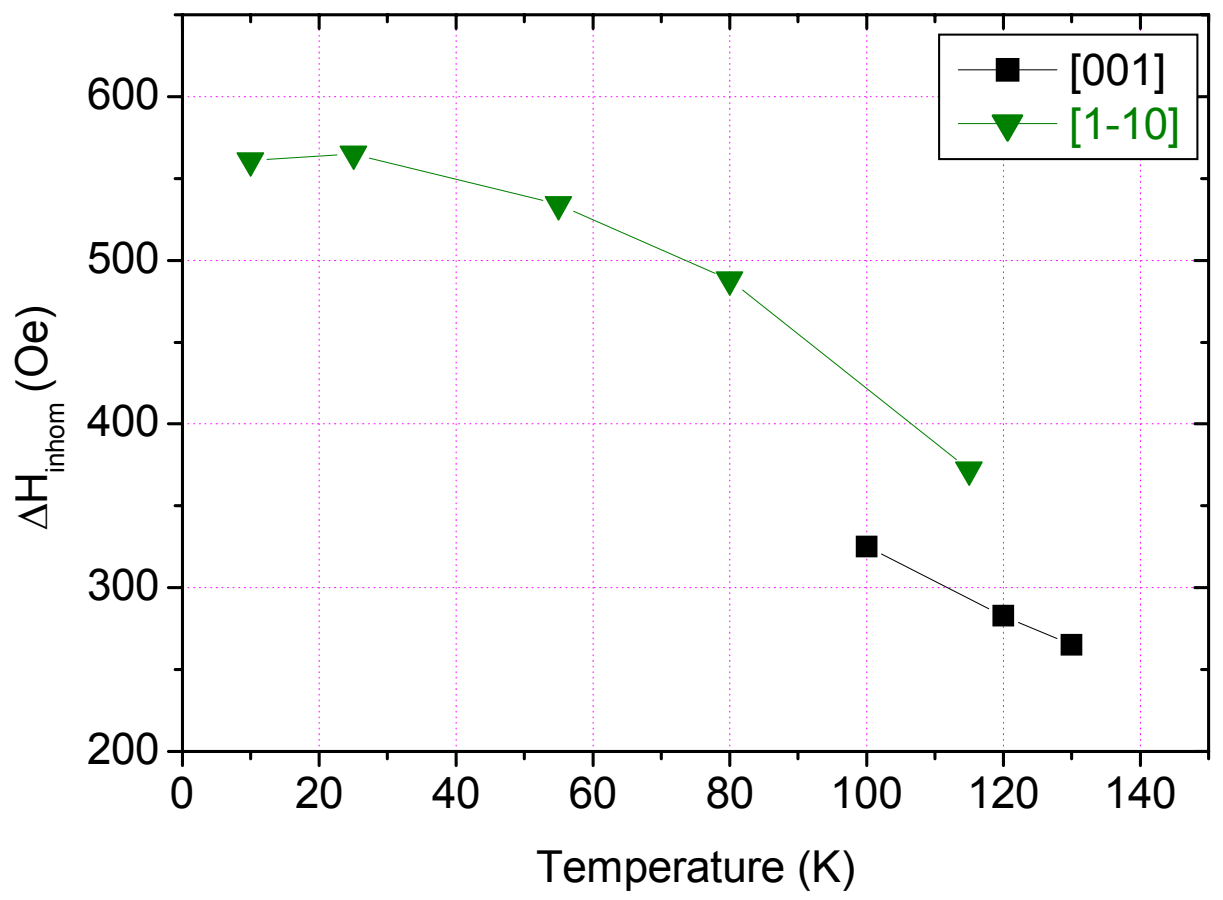

Figure 12

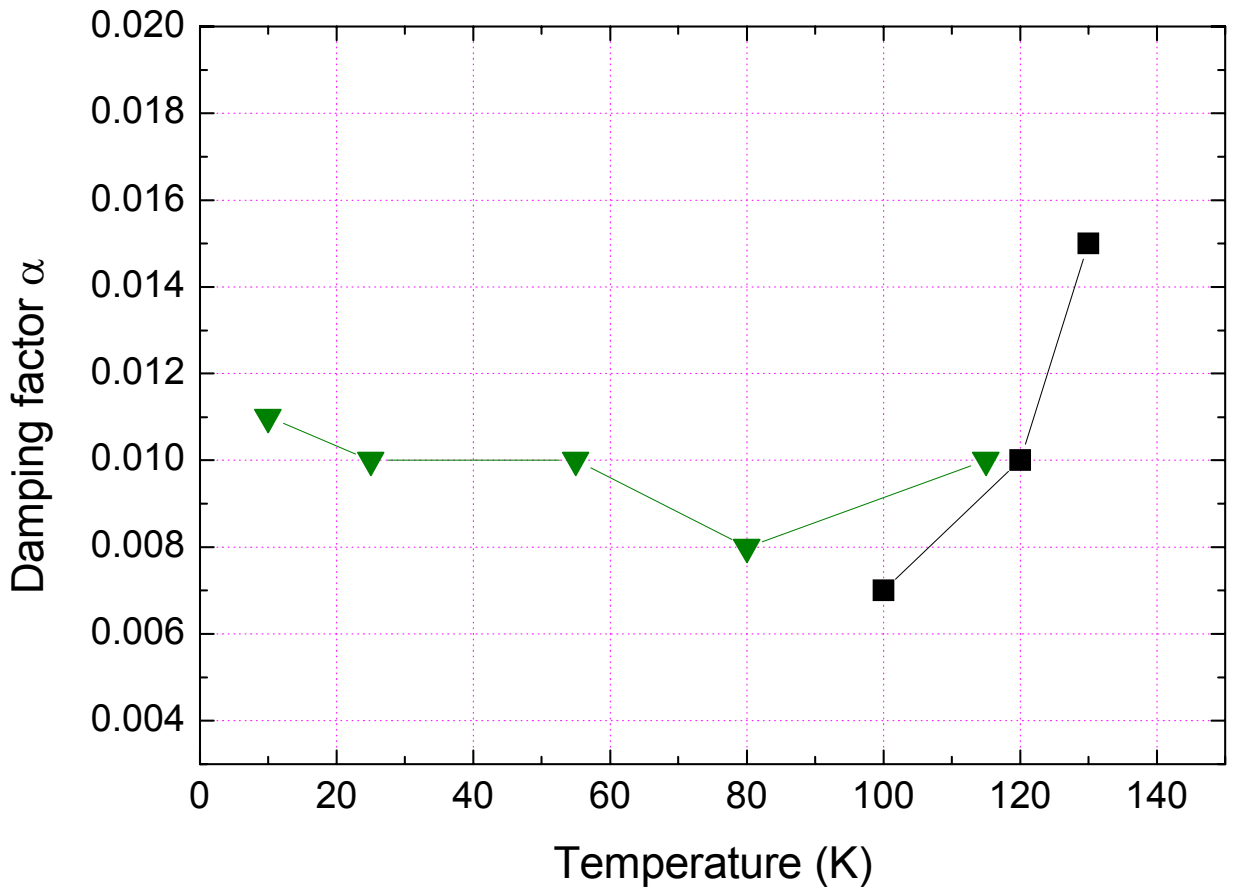

Figure 13